# Nanostructured and Modulated Low-Dimensional Systems

Albert Prodan, Herman J. P. van Midden, Erik Zupanič, Rok Žitko

Jožef Stefan Institute, Jamova 39, SI-1000 Ljubljana, Slovenia

albert.prodan@ijs.si

**Abstract**. Charge density wave (CDW) ordering in $NbSe_3$ and the structurally related quasi one-dimensional compounds is reconsidered. Since the modulated ground state is characterized by unstable nano-domains, the structural information obtained from diffraction experiments is to be supplemented by some additional information from a method, able to reveal details on a unit cell level. Low-temperature (LT) scanning tunneling microscopy (STM) can resolve both, the local atomic structure and the superimposed charge density modulation. It is shown that the established model for $NbSe_3$ with two incommensurate (IC) modes, $\boldsymbol{q_1}$ = (0,0.241,0) and $\boldsymbol{q_2}$ = (0.5,0.260,0.5), locked in at $T_1$=144K and $T_2$=59K and separately confined to two of the three available types of bi-capped trigonal prismatic (BCTP) columns, must be modified. The alternative explanation is based on the existence of modulated layered nano-domains and is in good accord with the available LT STM results. These confirm i.a. the presence of both IC modes above the lower CDW transition temperature. Two BCTP columns, belonging to a symmetry-related pair, are as a rule alternatively modulated by the two modes. Such pairs of columns are ordered into unstable layered nano-domains, whose $\boldsymbol{q_1}$ and $\boldsymbol{q_2}$ sub-layers are easily interchanged. The mutually interchangeable sections of the two unstable IC modes keep a temperature dependent long-range ordering. Both modes can formally be replaced by a single highly inharmonic long-period commensurate CDW.

## Introduction

The physical concepts for the stability of the charge density waves (CDW) in one-dimensional (1D) conductors were first given by Peierls [1], who predicted a semi-metallic modulated ground state. A small energy gap is opened at the Fermi level ($E_F$), accompanied by a loss of carriers and by anomalies in the electrical resistivity and the Pauli paramagnetism. The CDW is stabilized if the gain in electron energy more than compensates the increase in the elastic energy. Since $k_F$, the wave-vector at $E_F$, depends only on the filling of the conduction band, the CDW will in general be incommensurate (IC) with the underlying lattice. Large flat and approximately parallel sheets in the Fermi surface, separated by $\boldsymbol{q} = 2\boldsymbol{k_F}$ are responsible for a so-called nesting instability and the formation of a CDW.

An interesting property connected with CDWs is their sliding under the application of an external electric field. The effect is attributed to the ability of a CDW to overcome its pinning to the lattice and to the impurities [2]. The relatively small group of 1D compounds exhibiting such sliding includes in addition to the most thoroughly studied $NbSe_3$ and its isostructural monoclinic polymorph m-$TaS_3$ [3] also $NbS_3$ [4], $(TaSe_4)I$ [5], $(NbSe_4)_{10}I_3$ [6] and the "blue bronzes" $A_{0.3}MoO_3$ with A = K, Rb, Tl [7-10].

The monoclinic room-temperature (RT) basic structure of $NbSe_3$ ($a$ = 1.0006 nm, $b$ = 0.3478 nm, $c$ = 1.5626 nm, $\beta$ = 109.30°, space group $P2_1/m$) [11] is reproduced in Fig. 1. The unit cell contains three types of symmetry-related pairs of bi-capped trigonal prismatic (BCTP) columns with all type-I and type-III Nb-Se distances very similar (between 0.26 nm and 0.27 nm), while the inter-column Nb-Se bonds of the type-II BCTPs appear slightly longer (about 0.29 nm). Nb chains in Se cages form columns, aligned parallel to the monoclinic $\boldsymbol{b_0}$-direction. Strongly bonded corrugated layers, parallel to the $\boldsymbol{b_0}$-$\boldsymbol{c_0}$ monoclinic plane, are separated by wide Van der Waals (VdW) gaps (with Se-Se inter-column distances between 0.37 and 0.41 nm). The $NbSe_3$ basic structure is very specific:

- It is strongly anisotropic; in addition to its 1D nature, characterized by the BCTP columns, it has also a pronounced layered character.
- Three types of BCTP columns form the structure: the type-I (or orange-O) and type-III (or yellow-Y) columns are similar and with equilateral bases, while the bridging type-II (or red-R) columns appear with almost isosceles bases.
- The three column types are also very different with regard to their neighborhood in the unit cell: while all three appear in symmetry-related pairs, the type-III columns form continuous layers parallel to the $a_0$-$b_0$ planes, the type-I columns form well isolated pairs, and the type-II ones appear as isolated single columns.

It was argued from the very beginning [12] that two IC CDWs, which at least approximately add to a commensurate (CM) value, appear independently at different onset temperatures and are confined separately to two of the three available BCTP column types: $q_1$ = (0,0.241,0) below $T_1$ = 144 K to type-III and in addition $q_2$ = (0.5,0.260,0.5) below $T_2$ = 59 K to type-I columns. Such scenario appears to be in good accord with most experiments performed, including transport properties [13-22], electron [23] and X-ray diffraction studies [24-27], nuclear magnetic resonance (NMR) [28-30] and angularly resolved photoemission spectroscopy (ARPES) [31,32]. However, there are a few remaining open questions:

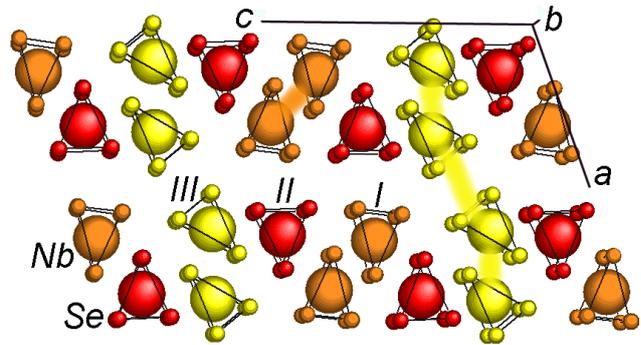

Fig. 1 The basic structure of NbSe$_3$ [11]. Large red (dark), orange (gray) and yellow (light) balls represent Nb atoms in type-II, type-I and type-III columns and the corresponding small balls the Se cages. Continuous layers of type-III and pairs of type-I columns are indicated.

- Why is CDW sliding a relatively rare phenomenon?
- Is it by accident that the IC components of the two $q$–vectors add within experimental error to a commensurate (CM) value, not only in NbSe$_3$, but also in the isostructural m-TaS$_3$?
- What is the origin of the "twinkling" domains, crossed by Moiré-like fringes, observed in satellite dark-field transmission electron microscopic (TEM) images [33]?
- Why are some important details, clearly revealed in the low-temperature (LT) scanning tunneling micsroscopic (STM) images [34-40] in apparent disaccord with expectations?

**The Domain Structure and the Alternative Model**

In an alternative approach [41] it is likewise presumed that only two of the three column types (III and I) are modulated. These form in the modulated ground state continuous sheets (type-III) and isolated pairs (type-I). However, individual symmetry-related pairs of columns are alternatively modulated by the two IC modes, which can formally be replaced by long-period (LP) CM values within $58b_0$: ($q_1$ = 0, 14/58, 0) and $q_2$ = (0.5, 15/58, 0.5). The two modes interchange easily between the two columns of a pair. The phase relationship between the two IC modes and the basic structure has no influence on the diffraction pattern (DP) and the combined modulation can formally be replaced by a single, highly inharmonic mode, obtained by beating between $q_1$ and $q_2$. The ordering of these basic units and its extents depend on temperature. With the disorder between the two modes taking place on a scale, small in comparison with the instrumental coherence regions, their contribution to the reciprocal space will depend on a convolution between the periodicities involved, i.e. the one of the basic structure and the combined inharmonic LP modulation. This can be easily verified by simple simulations of the DPs [42]. Since the columns are covalently bonded into layers, the two modes will be ordered into larger layered domains. Contrary, as a result of the very weak inter-layer bonding, the shortest dimensions of these domains will in average hardly exceed a few layers (Fig.2).

The domain model (Fig.2) is based on two predictions: first, at least one dimension of the domains is extremely small and second, the domains, though not in contact, remain over larger distances in proper phase relationships. The same number of columns, forming a certain domain, is modulated by the two modes, which as a rule appear in pairs. Such domains are unstable and can easily change their shapes with both modes appearing simultaneously. Since the modulation superstructure formed along the type-III columns forms sheets across the VdW gaps, it is stabilized at higher temperatures ($T_1$) than the one, which involves well separated type-I pairs ($T_2$). With the $q_1$ sections fully ordered along the type-III slabs parallel to the $a_0$-$b_0$ plane and with their two possible out-of-phase displacements along the type-I columns, the type-III columns will contribute only the $q_1$ satellites to the DP. The situation is reversed in case of $q_2$: the $q_2$ sections are separated by an even or an odd number of half shorter $q_1$ sections [25] along the type-III columns, but can be reordered along the type-I columns within an enlarged, body-centered modulation unit cell. Consequently, the type-III and type-I columns contribute only the $q_1$ and $q_2$ satellites, respectively to the DP [42].

**The Experimental Evidence**

A critical scattering was detected in case of NbSe$_3$ in X-ray diffraction experiments above both second order phase transitions [24-26]. The measured correlation lengths above the transitions were indeed small in comparison with the coherence regions, estimated to exceed hundred nanometers in synchrotron [43] and TEM experiments [44]. In an LT synchrotron radiation analysis [27] the atomic displacements along the three column types were refined with the original Wilson's model, but the final R-factors achieved for both satellite types were too large to discard beyond any doubt other possible solutions.

NMR measurements were performed at RT, 77 K and 4.2 K on powdered samples and on samples composed of aligned single crystals [28-30]. The RT $^{93}$Nb spectra with the nuclear spin quantum number I = 9/2, recorded with the magnetic field parallel to the crystallographic $b_0$ direction, clearly resolved 27 lines. One set of lines remained on cooling unchanged; the second became spread at 77 K and the third in addition at 4.2 K. This is as expected for the original model [12], where the Nb sites along the type-II columns are not modulated, while the $q_1$ mode is supposed to appear below $T_1$ along type-III and in addition the $q_2$ mode below $T_2$ along the remaining type-I columns. However, these results are also in accord with the domain model [41], since NMR is not able to distinguish between the Nb sites with regard to the weak CDW modulation.

Satellite dark-field TEM experiments performed on NbSe$_3$ [33] revealed above $T_2$ elongated strands, some 20 nm wide and 2000 nm long, crossed by unstable Moiré-like fringes of different widths (between 8 and a few 100 nm). These strands and fringes exhibited a characteristic "twinkling". While these features can hardly be explained with the old model, they are in accord with the new explanation.

LT STM results [34-38] are of greatest importance. Images of NbSe$_3$ reveal a few important details, which are in accord with both models:
1. At 77 K (i.e. between $T_1$ and $T_2$) only one surface column type per $c_0$ periodicity (type-III) is strongly modulated, while at 5K (i.e. below $T_2$) the number of such columns is doubled (type-III and type-I).
2. The CDW ordering at 5K enlarges the modulation unit cell into 2$c_0$.
However, some additional details are clearly in support of the domain model:
1. A doubled lateral periodicity into 2$c_0$ at 77K, detected either as a small difference in the intensity of the alternating strongly modulated (type-III) surface columns, or as a weak additional modulation along the second (type-I) type of columns are in support of the $q_2$ presence at that temperature.
2. Both (type-III and type-I) surface columns are at 5K modulated with the same mode and not alternatively with $q_1$ and $q_2$.

3. If subsurface columns are detected at 5K, they are either modulated by the alternative mode to the one detected at the surface, or both IC modes belonging to an overlapped surface/subsurface pair of columns shows a LP mode, obtained by beating between $q_1$ and $q_2$.

**Discussion**

X-ray crystallography remains beyond any doubt the most important technique for structural determination, but in case the order examined is extended only over sub-nanometer dimensions the results should be re-examined by some additional method [45,46]. All diffraction methods employed in nano-structured materials are faced with an important limitation, i.e. the relatively large irradiated samples.

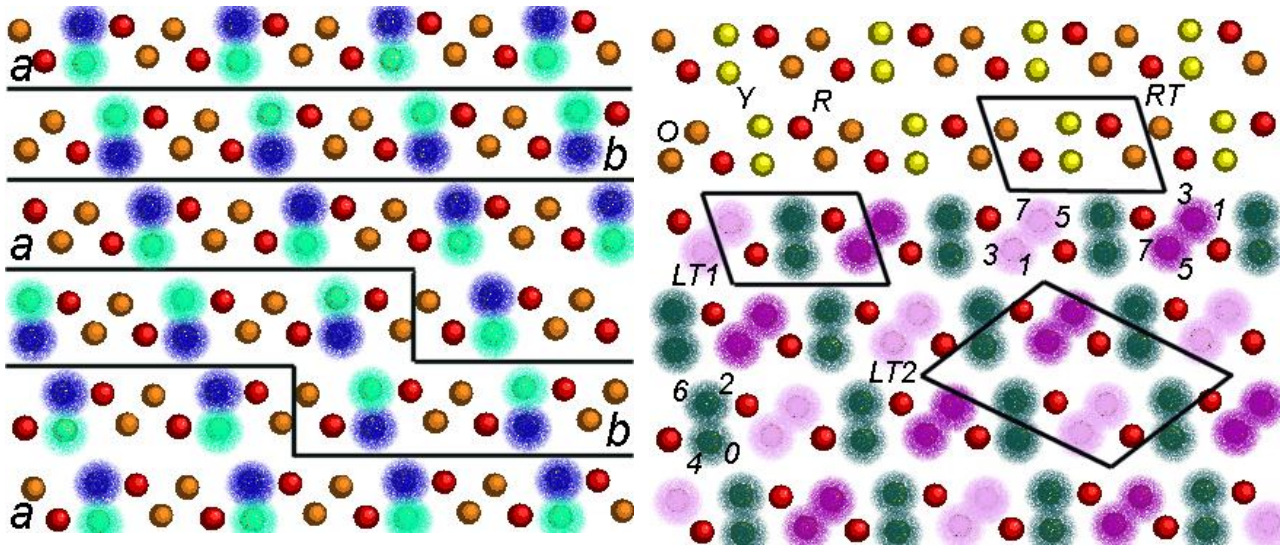

Fig. 2 A model with disordered $q_1$ (dark/blue) and $q_2$ (light/green) modes along type-III columns of NbSe$_3$ (with only Nb atoms shown) in the LT1 temperature region (i.e. between T$_1$ and T$_2$), showing cross-sections of nano-domains (left). Pairs of possible phase-shifts (in fractions of 1/8 of the 58$b_0$ periodicity) for the combined ($q_1 + q_2$) modulation along type-III (dark/marine) and type-I (dark&light/purple) pairs with the RT, LT1 and LT2 (i.e. below T$_2$) unit cell bases (right). The type-II columns are not modulated.

LT STM gives an alternative kind of structural information in comparison with the diffraction methods. It is the only method, where averaging does not represent a problem. It may suffer from other drawbacks; it is a surface sensitive method, the interaction between the scanning tip and the surface may induce changes, particularly if CDWs are involved, and subsurface contributions may complicate the obtained information. But the method is capable of revealing details in individual nano-domains, which may, due to their small sizes and due to averaging within experimental coherence regions, be overlooked by other methods.

Subsurface contributions to the LT STM images are of crucial importance in cases like NbSe$_3$ and the related structures. They reveal that two BCTP columns forming a symmetry-related pair are always modulated pair-wise, i.e. alternatively with the $q_1$ and $q_2$ modes. With both modes overlapped, STM shows the characteristic beating pattern [38].

A CDW ordering in NbSe$_3$ takes place on cooling along the type-III planes first (between T$_1$ and T$_2$) and in addition along the type-I pairs of columns (below T$_2$). However, this ordering is selective, with the $q_1$ and $q_2$ sections ordered along the type-III and type-I columns, respectively. This, together with the disorder on a nano-scale, which involves switching of the two modes between pairs of BCTP columns, results in the peculiar overall contribution to the reciprocal space. Although a particular mode ($q_2$ along type_III columns and $q_1$ along type-I columns) will give in a certain temperature region no detectable contribution to the DP, it will remain clearly visible in the corresponding STM images.

The origin of the unusual contribution to the reciprocal space of NbSe$_3$ is in the very specific layered nano-domains, formed parallel to the $b_0$-$c_0$ plane. These are composed of $q_1$ and $q_2$ sub-layers, which can be easily interchanged. The CDW sliding, characteristic for NbSe$_3$ and the related compounds is to be attributed to the easy switching between the $q_1$ and $q_2$ sub-layers of the unstable domains, which carry with their 14 and 15 charge maxima per 58 $b_0$ slightly different charges.

The two IC modes can formally be replaced by a single, LP CM modulation, obtained by beating between them and present along all columns, being modulated in a certain temperature region (type-III below T$_1$ and in addition type-I below T$_2$).

Finally, the origin of the unusual CDW ordering in NbSe$_3$ and the few related compounds is in their specific structures. These are composed of symmetry-related pairs of columns, which also compensate their common charge distribution pair-wise.

**Summary**

The two IC CDWs in NbSe$_3$ can be represented as LP CM modes with $q_1$ = (0,14/58,0) and $q_2$ = (0.5,15/58,0.5).

The entire reciprocal space of NbSe$_3$ can be reproduced with a single highly inharmonic modulation, obtained by beating between the two modes.

The explanation, based on the existence of highly anisotropic layered domains can be applied to the structurally related 1D compounds, e.g. m-TaS$_3$ with $q_1$ = (0,18/70,0) and $q_2$ = (0.5,17/70,0.5), whose structures are likewise characterized by symmetry-related pairs of columns.

**Acknowledgement**

Financial support of the Slovenian Research Agency (ARRS) is gratefully acknowledged.